# High-speed 4 × 4 silicon photonic electro-optic switch, operating at the 2 μm waveband

Jiawei Wang,[1] Jia Xu Brian Sia,[1,4,*] Xiang Li,[2] Xin Guo,[1] Wanjun Wang,[1] Zhongliang Qiao,[1] Callum G. Littlejohns,[3] Chongyang Liu,[2] Graham T. Reed,[3] Rusli,[1] and Hong Wang[1,*]

[1]*School of Electrical and Electronic Engineering, Nanyang Technological University, 50 Nanyang Avenue, Singapore 639798*
[2]*Temasek Laboratories, Nanyang Technological University, 50 Nanyang Avenue, Singapore 639798*
[3]*Optoelectronics Research Centre, University of Southampton, Southampton SO17 1BJ, UK*
[4]*Department of Materials Science & Engineering, Massachusetts Institute of Technology, Cambridge, MA, USA*
*\*jiaxubrian.sia@ntu.edu.sg/jxbsia@mit.edu*
*\*ewanghong@ntu.edu.sg*

**Abstract:** The escalating need for expansive data bandwidth, and the resulting capacity constraints of the single mode fiber (SMF) have positioned the 2-μm waveband as a prospective window for emerging applications in optical communication. This has initiated an ecosystem of silicon photonic components in the region driven by CMOS compatibility, low cost, high efficiency and potential for large-scale integration. In this study, we demonstrate a plasma dispersive, 4 × 4 electro-optic switch operating at the 2-μm waveband with the shortest switching times. The demonstrated switch operates across a 45-nm bandwidth, with 10-90% rise and 90-10% fall time of 1.78 ns and 3.02 ns respectively. In a 4 × 4 implementation, crosstalk below -15 dB and power consumption below 19.15 mW across all 16 ports are indicated. The result brings high-speed optical switching to the portfolio of devices at the promising waveband.

## 1. Introduction

With the spread of big data and 5G networks, optical communication systems are evolving to accommodate the expanding volume of information that needs to be transmitted over point-to-point or meshed networks [1-4]. Currently, the most commonly used wavelength for optical communication is at the C-band. However, there exists a theoretical maximum transmission capacity due to the Shannon limit [5,6]. To address the anticipated increase in capacity requirements, one promising approach is to initiate the development of a new waveband infrastructure. In regard to the abovementioned, the 2 μm waveband appears to be promising in view of the development of the low-loss hollow core photonic bandgap fiber (HC-PBGF) as well as the thulium-doped fiber amplifier (TDFA), critical aspects of a communication system [7-11]. HC-PBGF is an interesting media for optical transmission at 2 μm, indicating low optical nonlinearity, latency, transmission loss and high-power handling capabilities [12,13]. TDFA, on the other hand, enables high small-signal gain and low noise figures [14,15]. Due to high CMOS compatibility, silicon photonics has become a disruptive integrated optical platform which is easily scalable with low cost [16-20]. Over the years, substantial efforts have been dedicated to the development of integrated silicon photonic circuits operating at the 2 μm waveband. As a result, a comprehensive array of essential building blocks, including passive and active components such as waveguides [21,22], grating couplers [23,24], micro-ring resonators [25,26], multimode interferometers [27,28], lasers [29-31], photodetectors [32,33], and modulators [34,35] have been demonstrated. These advancements underscore the potential of silicon photonics at the 2 μm waveband as a new and prospective window for future development and applications, including optical communications, and while not limiting, biomedical monitoring, and sensing [36-39].

Among these key silicon photonics components, high-speed optical switches routing with low power consumption are crucial in optical communication [40], optical interconnection [41], and high-performance optical computing [42]. Photonic switching is enabled by the phase modulation of lightwaves, through effective refractive index manipulation in the waveguide arms. This modulation can be induced by the thermo-optic (TO) effect, where the phase shifter area is heated, or by the plasma dispersive effect, where free-carrier injection from PN/PIN junctions occurs upon the application of forward bias voltage to the device. Generally, TO switches offer simpler fabricating process [43,44], while plasma dispersive switches utilizing carrier injection, can achieve switching speeds in the nanosecond-scale: ~1000 times faster than TO switches. Furthermore, the thermal crosstalk inherent in TO switches precludes high integration densities. In the era of cloud computing, the Internet of Things (IoT), and artificial intelligence (AI), high-speed plasma dispersive switches are critical for efficiently managing a substantial volume of random data and responding to exploratory requests in real time, thereby enabling the ultra-fast switching of frequent short messages. Therefore, plasma dispersive switches become particularly crucial and are applicable to vast amounts of applications, including packet-switched multi-plane/multi-path wavelength division multiplexing (WDM) networks, low-latency interconnect between microprocessors, and high-throughput data center networks (DCN), where ultrafast switching speed on the order of nanoseconds or less is favored [45-48]. We note that the development of optical switches, specifically high-speed switch, at 2 μm have been lacking [49-54], while our group have reported a $1 \times 8$ optical switch [51]. Given the escalating demand for higher capacity and progressive trend towards photonics circuits at 2 μm [55], the development of high-speed switches becomes increasingly necessary and crucial.

In this work, we report a $4 \times 4$ electro-optic silicon switch operating at the 2-μm waveband based on the Mach-Zehnder interferometer (MZI) structure. Firstly, a $2 \times 2$ elementary cell is characterized. The power consumption for π-phase-shift ($P_\pi$) was measured to be 6.12 mW, extinction ratio (ER) higher than 16 dB and crosstalk (CT) lower than -16 dB are demonstrated across the 45-nm wavelength range (1968 - 2013 nm), considering both input ports. Subsequently, the elementary cell is scaled to a $4 \times 4$ optical switch by a waveguide crossing. Crosstalk of lower than -15 dB, with a minimum of -23 dB is characterized. Furthermore, dynamic characteristics of the $2 \times 2$ switch element are investigated, with 10-90% rise and fall time of 1.78 ns and 3.02 ns respectively. To the best of our knowledge, this is the highest switching speed in the waveband.

## 2. Device design

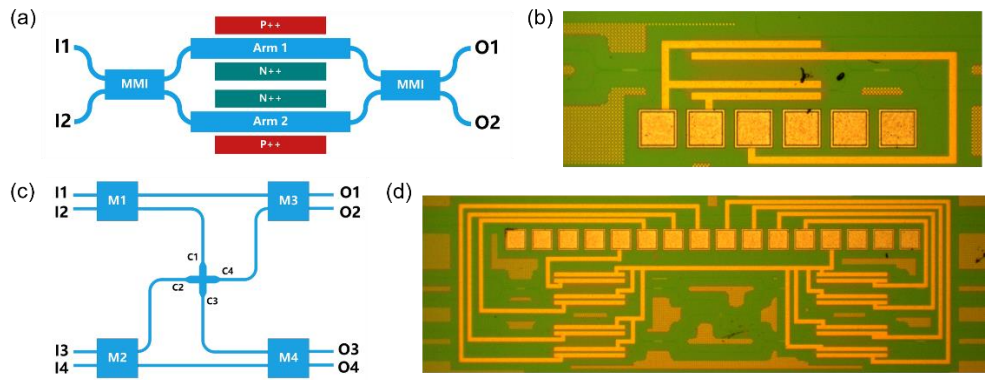

Fig. 1. (a) Schematic diagram and (b) micrograph image of the $2 \times 2$ elementary switches. (c) Schematic diagram and (d) micrograph image of the $4 \times 4$ optical switches.

The schematic diagram and micrograph view of the switches are shown in Fig. 1. The elementary switch cell (Fig. 1(a)-(b)) is based on a MZI structure featuring two 2×2 multimode interferometer (MMI) couplers shown in Fig. 2(a). The 2 × 2 MMI is designed using the Eigenmode expansion (EME) method, with a core region width ($W_0$) of 6 µm. The width of MMI input and output waveguides are tapered from 1.1 µm ($W_T$) to 0.6 µm (W) over a length of 10 µm ($L_T$), with a spacing of 2 µm. The optimal optical length of the MMI coupler was found to be 30.65 µm (Fig. 2(b)). The top-down electric field distribution of the designed MMI coupler is illustrated in Fig. 2 (c), where light is propagated from a single input, demonstrating efficient 3-dB splitting. The wavelength dependence of the coupler is indicated in Fig. 2(d), with a peak transmission of -0.57 dB across the simulated wavelength region.

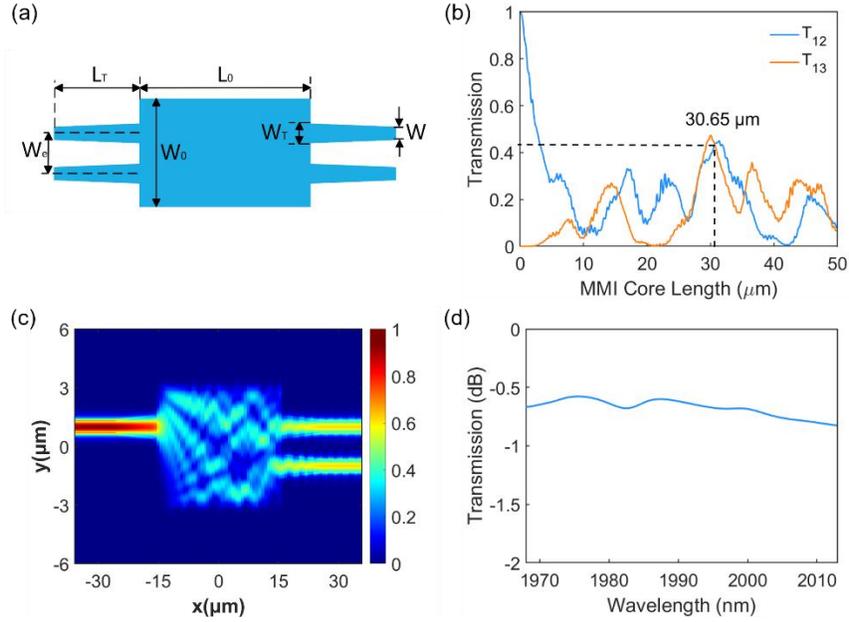

Fig. 2. (a) Schematic diagram of the 2 × 2 MMI. (b) 2 × 2 MMI coupler length optimization. (c) Electric-field distribution and, (d) transmission spectrum of the optimized 2 × 2 MMI.

Rib waveguides with 130 nm etch, are implemented for this work [56]. Eigenmode simulations are performed as a function of waveguide width to derive the single mode condition at 2 µm (Fig. 3(a)). Only the fundamental mode is supported when the width of the waveguide is smaller than 660 nm. As a result, the waveguide width was set to be 600 nm. The MZI structure includes two symmetric rib waveguide arms for the implementation of a phase shifter, with a PIN junction implemented on both arms via a 270-µm-long active PIN diode in consideration of switching and fabrication-induced phase error correction. The cross-section of the PIN junction is shown in Fig. 3(b). P++ (red) and N++ (green) regions are heavily doped to form ohmic contact with low contact resistance, positioned 1 µm away from the rib edge to minimize the free-carrier absorption loss. The electric field distribution of the optical mode is indicated at the inset of Fig. 3(b). Fig. 3(c) shows the carrier distribution at varying driving voltages. As the voltage increases, the concentration of electrons across the PIN section increases, leading to a change in the silicon material refractive index change, expressed as [57]

$$\Delta n = -\frac{e^2 \lambda^2}{8\pi^2 c^2 \varepsilon_0 n}\left(\frac{\Delta N_e}{m_{ce}^*} + \frac{\Delta N_h}{m_{ch}^*}\right) \qquad (1)$$

where $e$ is the electron charge, $\varepsilon_0$ is the permittivity of free space, $n$ is the inherent refractive index of crystalline Si, $\Delta N$ is the free-carrier concentration variation, and $m_c^*$ is the conductivity effective mass. Subscripts $e$ and $h$ represent electrons and holes, respectively. As a result, the effective refractive index decreases as illustrated in Fig. 3(d), thereby, a phase shift $\Delta\phi$ occurs as

$$\Delta\phi = 2\pi\Delta n_{eff} \frac{L}{\lambda} \qquad (2)$$

where $L$ is the length of the PIN phase shifter and $\lambda$ is the wavelength of the propagating mode.

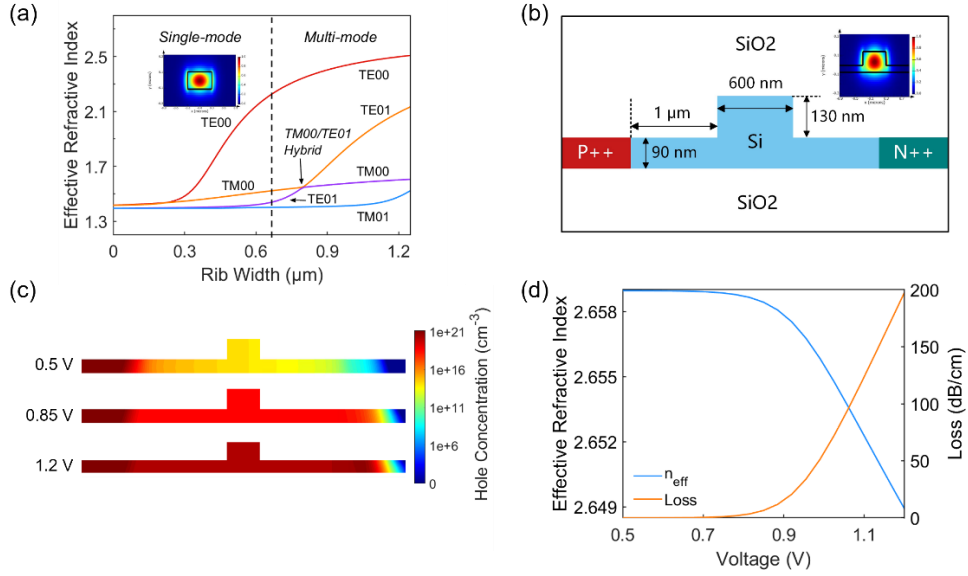

Fig. 3. (a) Single-mode condition of the strip waveguide at 2 µm wavelength range, subject to rib width. The inset shows the electric-field distribution of a 600-nm-width strip waveguide. (b) The schematic diagram of the cross-section of the slab waveguide; the red, green, and blue indicates the P++, N++ and intrinsic regions respectively. The inset shows the electric-field distribution of the fundamental optical mode. (c) Hole concentration distribution at varying driving voltage in the PIN junction region. (d) Effective refractive index and absorption loss of the PIN rib waveguide as a function of the driving voltage.

With regards to the elementary cell (Fig. 1(a)), a lightwave propagates to the input 2 × 2 MMI coupler and is split equally into the two arms. Through the application of a bias voltage to the MZI arms, free carriers are injected and the silicon material refractive index $\Delta n$ changes as indicated by Eq. (1), resulting in changes of the effective index of the propagating mode $\Delta n_{eff}$. A phase differential $\Delta\phi$ will be implemented across the MZI arms as indicated by Eq. (2). Subsequently, the lightwave from the two arms interferes at the output MMI forming an interferometric spectrum. The lightwave can be output from either two of the output port, subject to the phase differential applied across both arms.

To compensate for potential fabrication errors in the MZI arms, phase shifters are applied on both arms as mentioned earlier. Light from input port I1 (or I2) is designed to be routed initially to output port O1 (or O2), thereby achieving "bar" status via phase error correction. The forward voltage is then applied to the other arm to modulate the phase, redirecting the light to output port O2 (or O1). Upon reaching a π-phase-shift on the modulating arm, the switch shifts to the "cross" status.

The 4 × 4 switch design is derived from the elementary switch cell structure by connecting four 2 × 2 elementary switch cells through a waveguide crossing, as depicted in Fig. 1(c)-(d). As shown in Fig. 4(a), the dimensions of the waveguide crossing are optimized through inverse design which encompasses two orthogonal regions with core length $L_1$ = 18.4 μm, taper length $L_2$ = 5 μm and width $W_1$ = 3 μm, contributing to a total footprint of 28.4 μm × 28.4 μm. The electric-field distribution of the designed crossing is obtained by the Finite-Difference Time-Domain (FDTD) simulation as depicted in Fig. 4(b), with a simulated insertion loss and crosstalk of lower than 0.05 and -45 dB respectively.

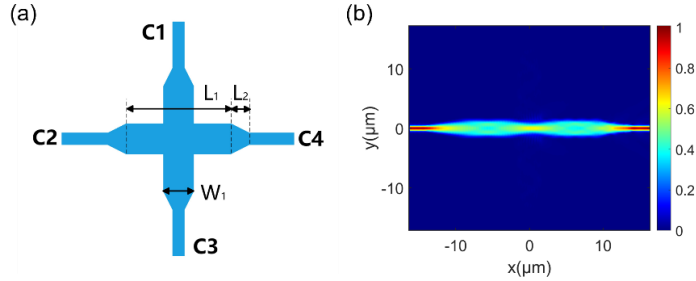

Fig. 4. (a) Schematic diagram and (b) electric-field distribution of the waveguide crossing.

## 3. 2 × 2 elementary switch characteristics

### 3.1 Static characteristics

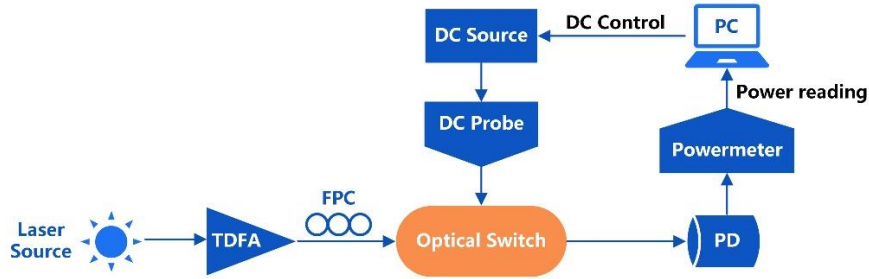

Fig. 5. Experimental setup for static characterization.

The experiment setup for evaluating static characteristics of the 2 × 2 switch elements is illustrated in Fig. 5. A wavelength (λ = 1990 nm) was input from a 2-μm tunable laser source (TLS) and subsequently amplified using a high-power TDFA, connected with a fiber polarization controller (FPC). Then the light was coupled in and out the switch device via edge coupling, facilitated by the lensed fiber and silicon inverse tapers (tip cross-section: 0.20 × 0.22 μm$^2$, length = 200 μm). The fiber coupling loss of -6.7 dB/facet was measured from a reference waveguide on the same chip. The output was subsequently directed towards a photodetector (PD). The device was subjected to a voltage bias applied via DC probes. Fig. 6 presents the measured IV curve of the fabricated PIN junction, with the turn-on voltage approximating 0.8 V.

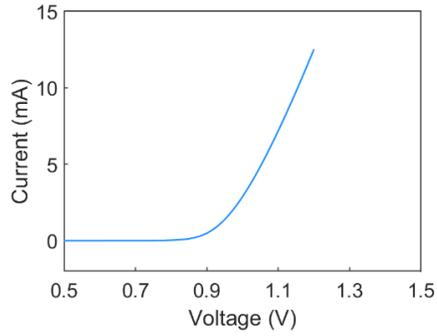

Fig. 6. Measured I-V curve of the PIN junction.

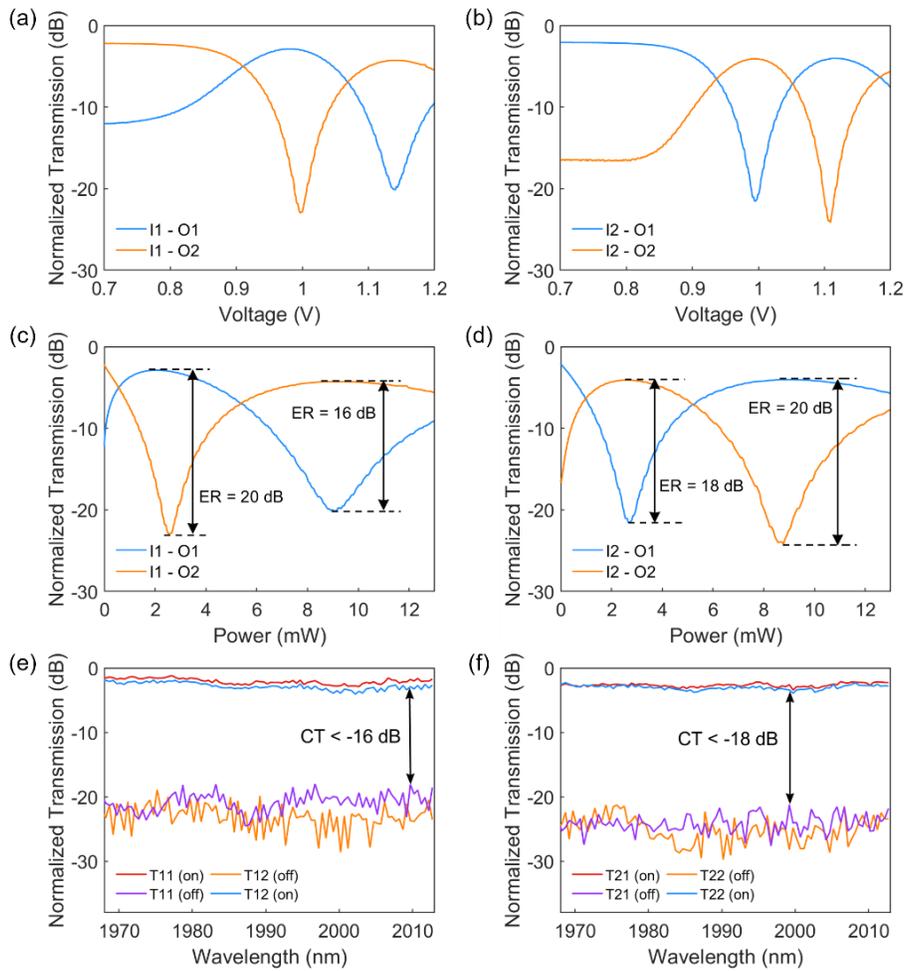

Fig. 7. Normalized transmission power as a function of the (a)-(b) applied voltage and (c)-(d) power for I1-input mode and I2-input mode. (e)-(f) Normalized transmission spectrum for I1-input mode and I2-input mode.

For the 2 × 2 elementary switch cell, I1 and I2 were designated as the input ports (Fig. 1(a)). A bias voltage of 0.92 V was applied to one arm to compensate the phase difference by device fabrication. With regards to I1 and I2, the output optical power (O1 and O2) as a function of bias voltage to the other phase shifter is indicated in Fig. 7(a) and (b) respectively. A $P_\pi$ of I1 and I2 are 6.56 (Fig. 7(c)) and 6.21 mW (Fig. 7(d)) respectively is obtained, with an ER of larger than 16 dB.

The wavelength dependance for both "bar" and "cross" statuses at the two output ports were measured and presented in Fig. 6(e)-(f). The insertion loss of $2.03 \pm 0.84$ dB and $2.57 \pm 0.65$ dB, as well as crosstalk lower than -16 dB and -18 dB, were recorded for I1-input mode and I2-input mode respectively, within a spectral bandwidth of 45 nm (1968 - 2013 nm). The plasma dispersive switch phase shifter operates in carrier injection mode, which accounts for the higher insertion loss due to free-carrier absorption (FCA) effects [58].

### 3.2 Dynamic characteristics

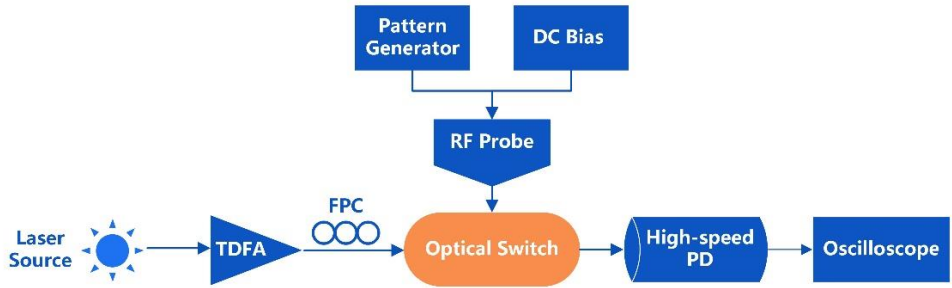

Fig. 8. Experimental setup for dynamic characterization.

The experimental setup for the dynamic characterization of the $2 \times 2$ elementary cell is shown in Fig. 8. A fixed laser source ($\lambda = 1990$ nm) was amplified by a TDFA and connected with the FPC before coupled in and out of the chip via lensed fiber-silicon inverse tapers. A RF signal with specific patterns was generated by the pattern generator (Anritsu MP1763C) and applied to the electro-optic phase shifters. The rise and fall time of the pattern generator are lower than 0.1 ns respectively (Fig. 9(a)). The signal transmitted from device's output ports was directed to a high-speed PD and subsequently linked with an oscilloscope (Agilent DSO93004L) to monitor the signal responding time.

The dynamic switching characteristics of the cell, subjected to a 50 kHz RF signal, 0.834 V peak-to-peak voltage ($V_{PP}$), biased ($V_{bias}$) at 0.35 V is shown in Fig. 9(b). The 10-90% rise (Fig. 9(c)) and fall time (Fig. 9(d)) are 1.78 and 3.02 ns respectively. To the best of our knowledge, this constitutes the fastest switching speed for silicon optical switches operating near 2 μm reported to date.

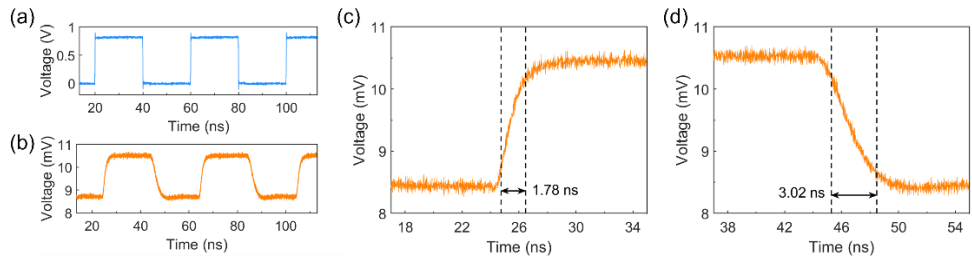

Fig. 9. (a) Applied 50 kHz RF signal with $V_{PP} = 0.834$ V and $V_{bias} = 0.35$ V. (b) The optical switch response with the (c) rising edge and (d) falling edge measured by the oscilloscope.

## 4. 4 × 4 switch characteristics

To further substantiate the scalability of the 2 × 2 switch element, we designed and fabricated the 4 × 4 electro-optic switch for the 2-μm waveband, which consists of four 2 × 2 switch elements (M1, M2, M3, and M4) through a waveguide crossing, as shown in Fig. 1(c)-(d). The measured transmission spectrum of the designed waveguide crossing (Fig. 4) is shown in Fig. 10, indicating a transmission loss of lower than 0.5 dB and a crosstalk of lower than -24 dB from 1968 nm to 2013 nm. With such configuration, 16 routing states were realized by four input ports (I1, I2, I3 and I4) and four output ports (O1, O2, O3 and O4). The measured power consumptions for both bar and cross states, and $P_\pi$ for each switch element, are detailed in Table 1; $P_\pi$ is obtained from the power differential between the cross and bar states. From the $P_\pi$ in Table 1, the average of 6.64 mW and standard deviation of 0.24 for all elementary cells are obtained, indicating a high degree of uniformity and scalability potential. Table 2 shows the minimum power consumptions for each optical switching path, where "1" represents "bar" status, "-1" represents "cross" status, and "0" represents the absence of input power in an ideal scenario. M1 - M4 (Fig. 1(c)) and C1 - C4 (Fig. 4(a)) refers to the labelling of the elementary cell and waveguide crossing output port respectively. The power consumption ranges from 5.31 to 19.15 mW.

The power spectrums at each output port for all 16 optical paths are shown in Fig. 11(a)-(p). The insertion loss for all 16 switch paths among the wavelength spectrum range was 5.47 ± 1.75 dB. As above mentioned, the insertion loss can be mainly attributed to the free-carrier absorption (FCA) effects. The crosstalk levels pertaining to Fig. 11 is summarized in Fig. 12, indicating maximum and minimum values of -15 and -23 dB from 1968 nm to 2013 nm (45-nm wavelength detuning range). The degradation of the switch performance is primarily due to the cascading of multiple elementary cells.

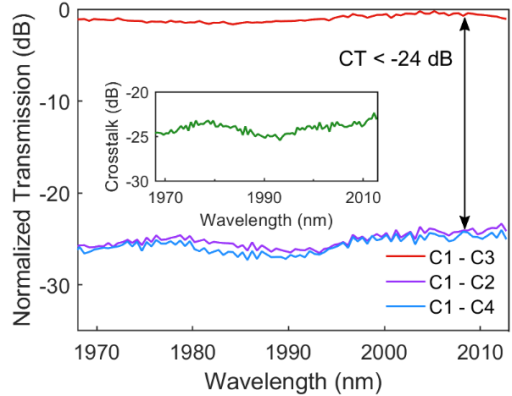

Fig. 10. Transmission spectrum of the waveguide crossing from port C1 to port C3. Inset: Crosstalk over the tested wavelength range.

Table 1. Power consumption of each elementary cell in the 4 × 4 switch[a]

| Switch Elements | M1 | M2 | M3 | M4 |
|---|---|---|---|---|
| Bar | 2.75 | 2.91 | 2.56 | 2.72 |
| Cross | 9.45 | 9.24 | 9.09 | 9.70 |
| $P_\pi$ | 6.70 | 6.33 | 6.53 | 6.98 |

[a]Values in Table 1 are in mW.

Table 2. Minimum power consumption for each optical path of the 4 × 4 switch[a]

| Optical Path | Switch Input Port - Output Port | M1 | M2 | Crossing Input Port - Output Port | M3 | M4 | Power (mW) |
|---|---|---|---|---|---|---|---|
| P1 | I1 - O1 | 1 | 0 | 0 | 1 | 0 | 5.31 |
| P2 | I1 - O2 | 1 | 0 | 0 | -1 | 0 | 11.84 |
| P3 | I1 - O3 | -1 | 0 | C1 - C3 | 0 | 1 | 12.17 |
| P4 | I1 - O4 | -1 | 0 | C1 - C3 | 0 | -1 | 19.15 |
| P5 | I2 - O1 | -1 | 0 | 0 | 1 | 0 | 12.01 |
| P6 | I2 - O2 | -1 | 0 | 0 | -1 | 0 | 18.54 |
| P7 | I2 - O3 | 1 | 0 | C1 - C3 | 0 | 1 | 5.47 |
| P8 | I2 - O4 | 1 | 0 | C1 - C3 | 0 | -1 | 12.45 |
| P9 | I3 - O1 | 0 | 1 | C2 - C4 | -1 | 0 | 12.00 |
| P10 | I3 - O2 | 0 | 1 | C2 - C4 | 1 | 0 | 5.47 |
| P11 | I3 - O3 | 0 | -1 | 0 | 0 | -1 | 18.94 |
| P12 | I3 - O4 | 0 | -1 | 0 | 0 | 1 | 11.96 |
| P13 | I4 - O1 | 0 | -1 | C2 - C4 | -1 | 0 | 18.33 |
| P14 | I4 - O2 | 0 | -1 | C2 - C4 | 1 | 0 | 11.80 |
| P15 | I4 - O3 | 0 | 1 | 0 | 0 | -1 | 12.61 |
| P16 | I4 - O4 | 0 | 1 | 0 | 0 | 1 | 5.63 |

[a] In Table 2, "1", "-1" and "0" represent "bar", "cross" and "no input" statuses separately in ideal scenarios.

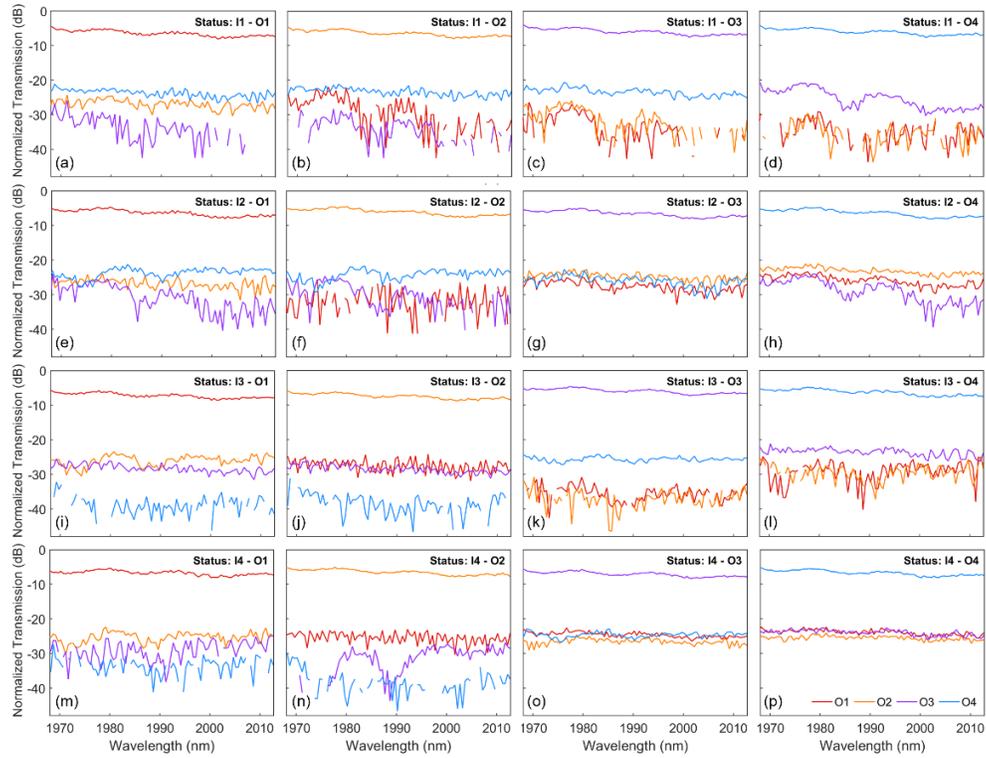

Fig. 11. Transmission spectrum of 4 × 4 switch for different routes. The partial absence of some lines is due to the power level that falls below the measurement threshold.

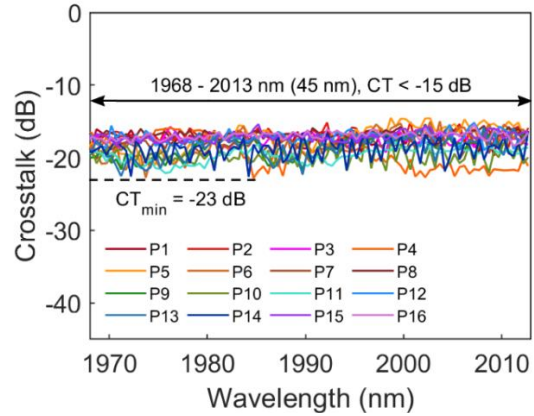

Fig. 12. Crosstalk levels of each output configuration in 4 × 4 switch.

Table 3 summarizes the key performance parameters of previously published silicon optical switches operating near 2 μm. Our designed 4 × 4 optical switch in this study exhibits favorable performance, including fastest switching speed, relatively low power consumption, and broad bandwidth.

**Table 3. Key Performance of Reported Silicon Optical Switches Near 2 μm**

| Working Mechanism | Device Structure | Configuration (Number of Cells) | Bandwidth (nm) | $P_\pi$ (mW) | CT (dB) | Rise/Fall Time (ns) | Year |
|---|---|---|---|---|---|---|---|
| TO | MZI | 2 × 2 (1) | 18 | 32.3 | N/A[a] | 15/15 (×10³) | 2019 [49] |
| TO | MZI | 1 × 1 (1) | 80 | 84.3 | N/A | 6/4 (×10³) | 2019 [50] |
| TO | MZI | 1 × 8 (7) | 75 | 23 | < -25 | 33.08/37.84 (×10³) | 2020 [51] |
| TO | MZI | 1 × 1 (2) | 100 | 19.2 | < -10 | 9.2/13.2 (×10³) | 2021 [52] |
| TO | MZI | 1 × 1 (1) | N/A | 25.21 | N/A | 3.49/3.46 (×10³) | 2021 [53] |
| TO | MRR | 1 × 1 (1) | N/A | 3.33 | N/A | 3.65/3.70 (×10³) | 2021 [53] |
| Plasma Dispersive | MRR | 2 × 2 (1) | N/A | 10.9 | N/A | 11.9/12.3 | 2022 [54] |
| Plasma Dispersive | MZI | 2 × 2 (1) | ~40 | 15.0 | -15.6 | 11.7/12.6 | 2022 [54] |
| Plasma Dispersive | MZI | 4 × 4 (4) | 45 | 6.21 | < -15 | 1.78/3.02 | 2023 (this work) |

[a]N/A represents that the value was not reported in the paper.

## 5. Conclusion

In this work, we present high-speed plasma dispersive switches operating at the 2 μm waveband. The static and dynamic characteristics of the devices were comprehensively evaluated. The 2 × 2 switch elements exhibited the fastest switching speed to date among silicon optical switches operating near 2 µm with rise and fall time of 1.78 ns and 3.02 ns respectively. The $P_\pi$ of 6.21 mW, with ER larger than 16 dB, and a crosstalk of lower than -16 dB across the 45-nm bandwidth were demonstrated with high uniformity at two input ports. The 4 × 4 switch indicates a minimum power consumption of 5.31 mW and a crosstalk of lower than -15 dB within the range of 1968 nm to 2013 nm for all 16 optical paths. Given its favorable performance and chip-scale size, the designed optical switch presents utility in promoting next-generation high-speed integrated applications at the 2 µm waveband.

**Funding.** This work was supported by the Ministry of Education (MOE) Singapore under Grant MOE-T2EP50121-0005.

**Disclosures.** The authors declare no conflicts of interest.

**Data availability.** Data underlying the results presented in this paper are not publicly available at this time but may be obtained from the authors upon reasonable request.